# Orientation Controlled Anisotropy in Single Crystals of Quasi-1D BaTiS$_3$

Boyang Zhao[1†], Md Shafkat Bin Hoque[2†], Gwan Yeong Jung[3], Hongyan Mei[4], Shantanu Singh[1], Guodong Ren[3], Milena Milich[2], Qinai Zhao[1], Nan Wang[1], Huandong Chen[1], Shanyuan Niu[1], Sang-Jun Lee[8], Cheng-Tai Kuo[8], Jun-Sik Lee[8], John A. Tomko[2], Han Wang[1,5], Mikhail Kats[4], Rohan Mishra[3], Patrick E Hopkins[2,6,7] and J. Ravichandran[1,5*]

[1]Mork Family Department of Chemical Engineering and Materials Science, University of Southern California, Los Angeles, California, 90089, USA

[2] Department of Mechanical and Aerospace Engineering, University of Virginia, Charlottesville, Virginia, 22807, USA

[3] Department of Mechanical Engineering and Material Science, Washington University in St. Louis, St. Louis, MO, 63130, USA

[4]Department of Electrical and Computer Engineering, University of Wisconsin–Madison, Madison, WI, 53706, USA.

[5]Ming Hsieh Department of Electrical Engineering, University of Southern California, Los Angeles, California, 90089, USA

[6] Department of Materials Science and Engineering, University of Virginia, Charlottesville, Virginia, 22904, USA

[7] Department of physics, University of Virginia, Charlottesville, Virginia, 22904, USA

[8] Stanford Synchrotron Radiation Lightsource, SLAC National Accelerator Laboratory, Menlo Park, CA 94025, USA.

[†] These authors contributed equally: Boyang Zhao, Md Shafkat Bin Hoque.

[*] j.ravichandran@usc.edu

Low-dimensional materials with chain-like (one-dimensional) or layered (two-dimensional) structures are of significant interest due to their anisotropic electrical, optical, thermal properties. One material with chain-like structure, BaTiS$_3$ (BTS), was recently shown to possess giant in-plane optical anisotropy and glass-like thermal conductivity. To understand the origin of these effects, it is necessary to fully characterize the optical, thermal, and electronic anisotropy of BTS. To this end, BTS crystals with different orientations ($a$- and $c$-axis orientations) were grown by chemical vapor transport. X-ray absorption spectroscopy (XAS) was used to characterize the local structure and electronic anisotropy of BTS. Fourier transform infrared (FTIR) reflection/transmission spectra show a large in-plane optical anisotropy in the $a$-oriented crystals, while the $c$-axis oriented crystals were nearly isotropic in-plane. BTS platelet crystals are promising uniaxial materials for IR optics with their optic axis parallel to the $c$-axis. The thermal conductivity measurements revealed a thermal anisotropy of ~4.5 between the $c$- and $a$-axis. Time-domain Brillouin scattering showed that the longitudinal sound speed along the two axes is nearly the same suggesting that the thermal anisotropy is a result of different phonon scattering rates.

# I. INTRODUCTION

Lower dimensional materials, both natural and synthetic, have been widely studied for the anisotropy in their electronic and optoelectronic properties.[1–4] Recently, there is a growing interest in materials with large anisotropy in optical and thermal properties. Optical anisotropy is widely used for quasi-phase matching in nonlinear optical media,[5] and in optical elements such as polarizers, wave plates, and filters.[6–8] Large thermal anisotropy is desirable for controlling heat flow in a broad range of applications ranging from electronics to acoustics.[9,10] Although the technological drivers for such anisotropic materials are evident, the engineering design of these technological applications using largely isotropic materials is demanding. Hence, the key bottleneck is the lack of suitable materials with the desired anisotropy.

The degree of anisotropy of the physical properties depends on the structural and chemical anisotropy of the material. Two-dimensional (2D) layered materials, such as graphene (graphite),[1] $MoS_2$,[3] and phosphorene (black phosphorus or BP),[4] have been studied extensively for the anisotropy between their out-of-plane and in-plane properties. Quasi-1D crystals possess 1D chains as building blocks and offer uniaxial anisotropy, which can drive novel electronic properties such as metal-insulator transitions and superconductivity.[11] Specifically, quasi-1D materials, such as $TiS_3$, have been studied for their in-plane optical and thermal anisotropy parallel and perpendicular to the Ti chains.[12,13]

A recent addition to the family of quasi-1D materials is $BaTiS_3$ (BTS). BTS is reported to crystallize in the hexagonal $P6_3mc$ $BaNiO_3$-type structure whose face sharing $TiS_6$ octahedra forms quasi-1D chains along the *c*-axis.[14,15] The large difference in the electronic polarizability between the inter- and intra-chain directions, combined with the inherent structural anisotropy give rise to strong optical anisotropy. Niu *et al.* (2018) reported[15] that BTS single crystals demonstrate a giant optical anisotropy with difference in the real part of the refractive index along different axes of up to 0.76 in the mid- to long-wave infrared regime, and a strongly dichroic window of 1.5–4.5 µm. Sun *et al.* (2020)[16] further demonstrated that BTS single crystals possess glass-like ultralow thermal conductivity along the [100] direction (*a*-axis), and showed evidences for enhanced phonon scattering mediated by tunneling of Ti ions between degenerate double-well potential[16]. However, full characterization of the optical, thermal, and electronic properties of BTS along different crystallographic directions remains a challenge, as single crystals with different orientations have not been available.

Here, we report the synthesis of BTS crystals with platelet morphology in two different orientations, (100) and (001), aside from the majority needle-like crystals using chemical vapor transport (CVT). We verified the presence of a uniaxial optical axis parallel to the $TiS_6$ chains (*c*-axis) *via* Fourier transform infrared (FTIR) spectroscopy and a giant optical anisotropy between the directions parallel and perpendicular to the *c*-axis consistent with past reports.[15,17] We performed time-domain thermoreflectance (TDTR)[18,19] measurements at room temperature and found the thermal conductivity values to be 1.69 ± 0.24 and 0.38 ± 0.06 $Wm^{-1}K^{-1}$ along the *c*- and *a*-axis, respectively. Furthermore, glass-like temperature dependence is observed along both directions, as seen in amorphous materials. The thermal anisotropy is likely caused by the differences in phonon scattering rates along the two directions as we find the longitudinal sound speed measured via time-domain Brillouin scattering (TDBS) to be nearly the same along the two crystallographic directions.

## II. Crystal Growth and Mechanism

Figure 1a and 1b show a schematic projected view of the crystal structure of BaTiS$_3$ (BTS) for (100) or (010), and (001) crystal orientations respectively. The orientation of the crystal is defined by the orientation of the as-shown projected surface. Our past growth efforts have resulted in both needle-like and platelet-like crystals with (100) orientation.[15] In this work, we show that (001)-oriented BTS crystals can be also synthesized using vapor transport growth. A detailed account of the growth methods is available in the Methods section.

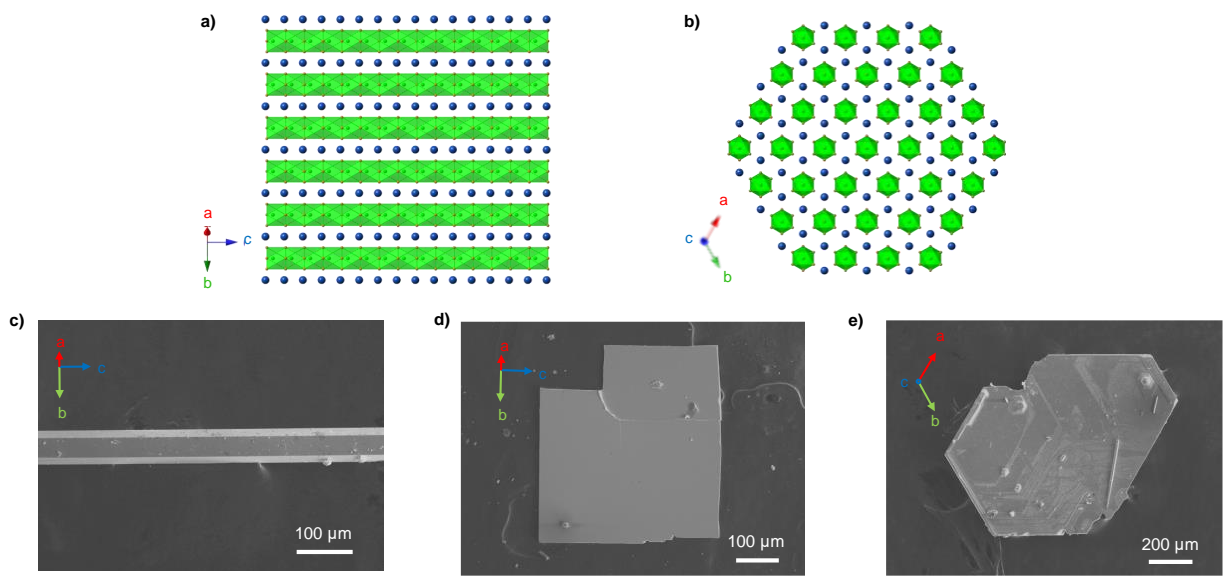

**Figure 1.** BTS crystals with different morphologies and orientations: (a) Projected view of the crystal structure of (100)- or (010)-oriented BTS visualizing quasi-1D TiS$_3^{2-}$ chains along the *c*-axis; (b) Projected view of the crystal structure of (001)-oriented BTS; Scanning electron microscopy (SEM) images and crystal orientations of (c) BTS needle, (d) (100)-oriented BTS platelet, and (e) (001)-oriented BTS platelet. Crystal shape and surface terraces can be used to distinguish them.

Figure 1 c, d, and e show the SEM image of BTS needle, and (100)- and (001)-oriented BTS platelets, respectively. Compared to the needle and rectangular platelet-shaped BTS (100) crystals, BTS (001) crystals show distinct hexagonal faceting, suggesting hexagonal symmetry for this face. The typical thickness of the platelets was ~ 5-20 μm, while needles were ~ 20-50 μm thick. Aside from the crystal morphology, we used X-ray diffraction (XRD) to determine the crystalline orientation (Figure S2 c, d, e). Extinct reflections of 00*l*, where $l = 2n + 1$, is a distinct sign of (001) orientation. The measured rocking curves show narrow full-width-at-half-maximum (~0.011 – 0.013°) for all BTS morphologies.

We performed a systematic investigation of the effect of growth parameters on the orientation and size of the crystals (Table I). We observed a sensitive dependence of the orientation of the crystals on the growth temperature, suggesting a subtle temperature-dependence of the surface energy for the different facets. Further, the surfaces that possess large surface energies (enthalpic component) tend to be more stable at higher temperatures as the entropy plays a more dominant role. Based on this argument, a temperature below 1020°C is too low to strike a balance between nucleation and growth to form large-area crystalline surfaces, thus making polycrystalline powders the dominant product. As the growth temperature is raised above 1020°C, the (100) surface becomes most favored amongst all the terminations, and the (100) orientation of BTS crystal nuclei becomes more stable and sustains appreciable growth rate, forming BTS needles. When the temperature goes up above 1055°C, the (001) orientation is also favored. Between 1020°C and 1055°C, certain amount of growth along <001> is allowed to form a larger (100) surface, enabling BTS platelets with (100) orientation. We usually find larger crystals at higher temperatures, where crystals of both morphologies coexist, but sintering is found to be dominant at temperatures higher than 1060 °C where free-standing crystals were not observed.

Table I. BTS of Different Shapes and Orientations

| Crystal | Reaction Temperature | Crystal Morphology | Surface Orientation |
|---|---|---|---|
| BTS needle | 1020 – 1060 °C | Needle-like | {100} |
| (100)-oriented BTS | 1040 – 1060 °C | Rectangular platelet | (100) |
| (001)-oriented BTS | 1055 – 1060 °C | Hexagonal platelet | (001) |

In-plane orientations of BTS (100) and BTS (001) were then determined by XRD pole figure analysis (See Supporting Information II), by studying the rotational symmetry of the easiest accessible reflections. Figure 1 c, d, e and Figure S1 a, b, c show the SEM and optical images of the crystals with the three different morphologies and the measured orientation noted by the axes-legend. The symmetry matches with the reported BTS structure with no evidence for a non-stoichiometry induced incommensurate structure (See Supporting Information III).[20]

### III. Polarization Dependence of X-ray Absorption Spectroscopy

X-ray absorption spectroscopy (XAS) is widely used to determine the local symmetry and electronic structure of materials.[21,22] Figure 2a shows the X-ray absorption near the Ti $L_{2,3}$ edge for BTS (100) and BTS (001) (noted as blue and orange colored spectra respectively), which was obtained in the partial florescence yield (PFY) mode. Due to the crystal field of TiS$_6$ octahedra, the five-fold degenerate $3d$ orbitals of Ti are split into a triply degenerate $t_{2g}$ and doubly degenerate $e_g$ orbitals, labeled as A, B at the $L_3$-edge (i.e., $2p_{j=3/2} \to 3d$ dipole transition) and C, D at the $L_2$-edge ($2p_{j=1/2} \to 3d$). The crystal field splitting, which corresponds to the local symmetry, ($E_{e_g} - E_{t_{2g}}$) of BTS (1.8 eV) turns out to be much smaller than TiS$_2$ (2.1 eV).[23] It implies a distorted coordination for the TiS$_6$ octahedra, which also induces additional multiplet states B' and B" (D") that appears as shoulders of $e_g$ in the XAS spectra at $L_2$ ($L_3$)-edge. These experimental spectral features, including the multiplet states, are qualitatively consistent with the reported structure. We conducted first-principles density functional theory (DFT) calculations for $L_3$-edge and constructed $L_2$ using the experimental $L_2$ - $L_3$ splitting of 5.7 eV. Figure 2a presents the Ti $L_{2,3}$ XAS spectrum calculated by first-principles DFT calculations (noted as simulated BTS), showing multiplet states matching well with the experimental spectral features for $L_3$-edge between 456 and 460 eV, A, C and B.

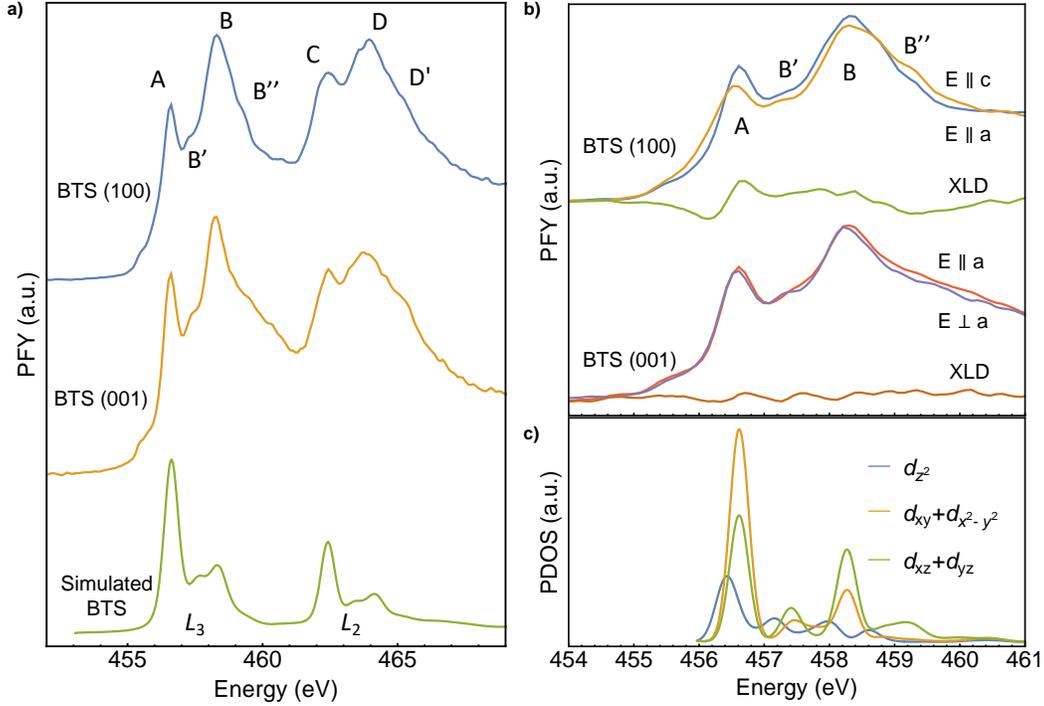

**Figure 2.** Ti $L_{2,3}$ XAS spectra of BTS. (a) XAS of BTS (100), BTS (001), along with BTS's DFT-calculated XAS spectrum. Calculated spectra are shown with a 0.2 eV Gaussian broadening. (b) Normalized linearly polarized XAS spectra of BTS (100) and BTS (001) along the principal axes. Polarizations are aligned with the primary axes of the crystals. XLD represents the difference of the XAS signals (or spectra) for orthogonal polarizations. (c) Calculated PDOS of Ti $3d$ of BTS. The degenerate ($d_{x^2-y^2}$, $d_{xy}$) and ($d_{xz}$, $d_{yz}$) orbitals are projected by summation of each component.

To better understand the local structure (*i.e.*, anisotropy of Ti), we carried out polarization-dependent XAS measurement. Figure 2b shows the linearly polarized XAS spectra of BTS (100) and BTS (001) crystals with linearly vertical (LV) and linearly horizontal (LH) polarizations. Note that the *a*-axis was aligned parallel to LV polarization for both crystal orientations (representing **E** // a), and the polarization dependence, *i.e.,* x-ray linear dichroism (XLD), is the difference between XAS spectra measured using the LV and LH polarizations. For BTS (100), the polarization dependence of XAS is remarkable. For the ***E*** // *c* polarization, the overall spectral features are weaker and broader. In particular, this broader feature is most obvious in A. We also observe clear shifts in both the absorption features, A and B, giving rise to an increase in the crystal-field splitting from 1.79 eV (***E*** // *a*) to 1.84 eV (***E*** // *c*). Such anomalous polarization-dependent crystal-field splitting gap has also been reported in $TiS_2$, whose local electronic structure is strongly affected by the high metal-ligand hybridization and significant ligand field.[24] BTS (001), by comparison, does not show a substantial linear dichroism, verifying a more isotropic local structure. To understand this behavior, we analyze the orbital projected density of states (PDOS) of BTS obtained from DFT calculations.

Figure 2c shows the resulting PDOS of Ti *d*-states. Overall, it shows qualitative agreement with the simulated XAS spectra, showing three main peaks at the energy of A, B and C. Ti *d*-states at the $L_3$ edge are a linear combination of orbitals $d_{z^2}$, $d_{x^2-y^2}$, $d_{xy}$, $d_{xz}$ and $d_{yz}$, where *x*, *y* and *z* are defined in Cartesian coordinate system (*x* ∥ *a, z* ∥ *c*, then *y* ⊥ *a*). Herein, $d_{x^2-y^2}$ and $d_{xy}$ or $d_{xz}$ and $d_{yz}$ orbitals are doubly degenerate. Under the trigonal crystal field with face-shared TiS$_6$ octahedra, $t_{2g}$ level can further split into a singlet $a_{1g}$ and doublet $e_g^\pi$ levels, consisting of solely $d_{z^2}$ orbital and linear combinations of $d_{x^2-y^2}$, $d_{xy}$, $d_{xz}$, and $d_{yz}$ orbitals, respectively.[25,26] The low-lying $d_{z^2}$ bands are highly overlapped with the other $d$-bands, suggesting that the $t_{2g}$ splitting into $a_{1g}$ and $e_g^\pi$ is too weak to be observed in the experiment. The relative position and DOS distribution of these orbitals determine the polarization dependence. At the absorption feature A, $d_{z^2}$ has the lowest energy, bringing A to a much lower energy for ***E*** // *c*. The absorption feature B is composed of excitations to ($d_{x^2-y^2}$, $d_{xy}$) and ($d_{xz}$, $d_{yz}$) orbitals at 458.3 eV but $d_{z^2}$ contributed to the ***E*** // *c* shoulder at 458.6 eV. The multiplet absorption peak B' is composed of both degenerate ($d_{x^2-y^2}$, $d_{xy}$) and ($d_{xz}$, $d_{yz}$) orbitals while B" arises from only ($d_{xz}$, $d_{yz}$) so B" shows stronger polarization dependent. Although negligible in the PFY mode, unknown surface oxides (Supplemental Information IV) contribute to noise level in the XLD feature near 460 eV. This model and corresponding interpretation thus captures all the observed features of the local anisotropy of quasi-1D BTS.

### IV. Optical Anisotropy

We discuss the optical anisotropy of BTS, specifically in the context of the newly available (001) oriented platelets. Figure 3a and Figure 3b show the normal-incidence transmittance and reflectance spectra of BTS (001) crystal platelets with in-plane polarization of 0°, 45° and 90° with respect to the *a*-axis of the crystal. Fabry–Pérot fringes indicate sufficient smoothness of the surfaces to achieve interference between the two surfaces. The near-identical spectrum for all polarizations suggests a highly isotropic crystal face. The absorption edge is observed at 0.76 eV, while the anomalous spike at 0.29 eV comes from atmospheric absorption of $CO_2$. This agrees well with the spectra we originally reported for (100)-oriented BTS,[15] but along an opposite axis. In the spectra shown in Figure 3a, b, the electric field is always perpendicular to the *c*-axis. This result is a clear evidence that the polarization assignment of our first published optical properties of BTS crystals of (100)-oriented BTS was flipped, as noted in our recent erratum.[27] Figure 3c is the polar map of polarization-dependent transmittance and reflectance at a wavelength of 2 μm. The optical axis is perpendicular to BTS (001), and hence, the axes perpendicular to the *c*-axis are ordinary axes. Such isotropic transmission windows extend to the longest wavelength measured here (~16 μm), making BTS (001) a candidate for IR-transparent optical elements.

In contrast to (001)-oriented BTS, (100)-oriented BTS is optically anisotropic in-plane. Transmission and reflectance spectra of (100)-oriented BTS platelet with in-plane polarization at 0°, 30°, 60°, and 90° with respect to the *c*-axis (Figure 3d, e) show an obvious polarization-dependent IR transmittance. Transmittance spectra as a superposition of the previously reported ordinary and extraordinary transmittance shoulder at 0.76 eV and 0.27 eV[15] are observed. But the highest transmittance is observed at polarization normal to *c*-axis, verifying the corrected polarization assignment.[27] Consistent with the transmittance shoulders, Fabry–Pérot fringes also disappear at this energy. The transmittance and reflectance values at a wavelength of 2 μm are plotted as a function of polarization orientation in a polar plot in Figure 3f, with large anisotropy in both reflectance and transmittance. The transmittance anisotropy between 0.4 eV and 0.7 eV makes BTS highly dichroic, while reflectance anisotropy verified BTS to be birefringent crystals. We performed unpolarized FTIR spectroscopy on multiple BTS platelets and needles grown at different times to confirm the reproducibility of these results. (Supporting information V).

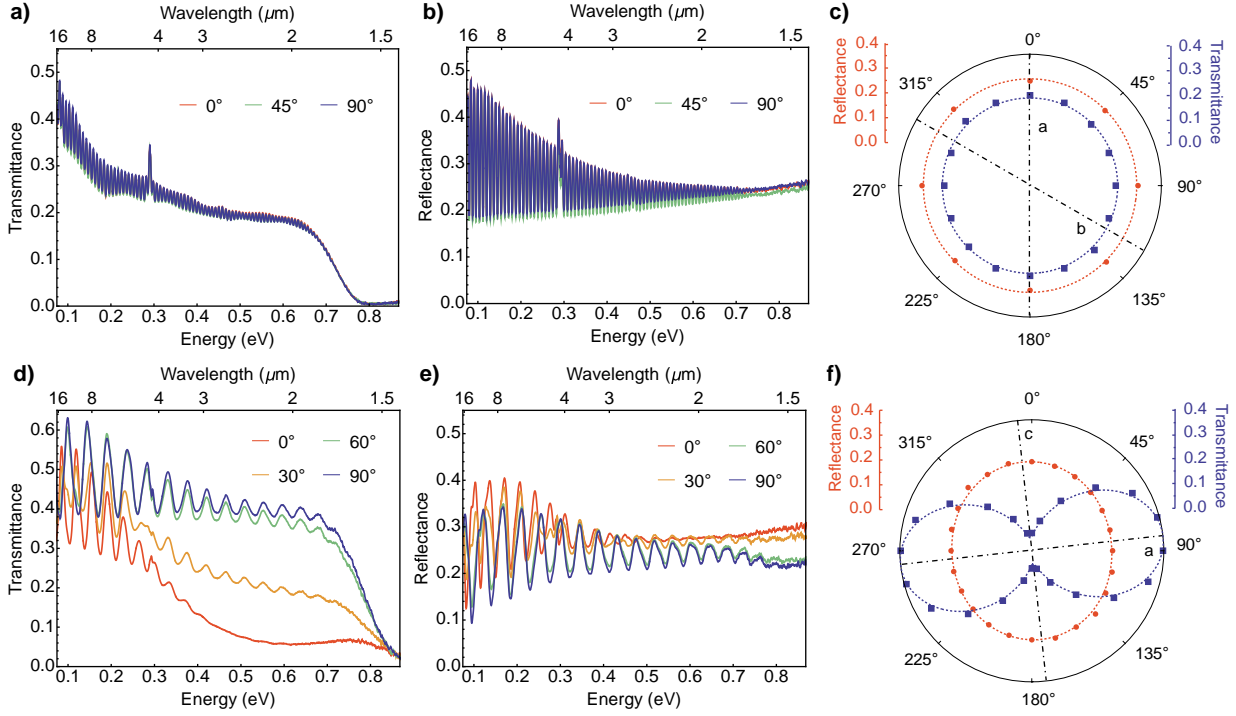

**Figure 3.** Optical spectroscopy of BTS platelet crystals by polarization-resolved FTIR. (a) Transmittance and (b) Reflectance spectra of BTS (001) crystal with polarization at 0°, 45° and 90° with respect to the *a*-axis. (c) Polar plot of transmittance and reflectance values of BTS (001) at 2 µm. (d) Transmittance and (e) Reflectance spectra of a (100) crystal with polarization at 0°, 30°, 60° and 90° with respect to the *c*-axis. (f) Polar plot of transmittance and reflectance values of BTS (100) at 2 µm. Transmittance plot shows a "dumbbell" shape and reflectance plot is elliptical. When fitted to second-order sinusoidal relationship to polarization, ordinary (84°) and extraordinary (354°) axes matches with the *a*-axis and *c*-axis of the crystal.

## V. Thermal Anisotropy

The growth of BTS along different crystallographic orientations allows us to study the anisotropy in its thermal transport. To measure thermal transport anisotropy, we use the time-domain thermoreflectance (TDTR).[28,29] Figure 4a shows the thermal conductivity along the [001] direction (*c*-axis, $k_{\parallel}$) measured in this study as well as the measurements of Sun *et al.*[16] along the [100] direction ($k_{\perp}$). We verified the thermal conductivity along the [100] crystallographic direction near room temperature (Figure 4a) to be in agreement with that of Sun *et al.*[16] The thermal conductivity of BTS along the [001] and [100] directions ($k_{\parallel}$ and $k_{\perp}$, respectively) were measured to be $1.69 \pm 0.24$ and $0.38 \pm 0.06 \, W \cdot m^{-1} \cdot K^{-1}$, which gives rise to an anisotropy ratio ($k_{\parallel} / k_{\perp}$) of ~4.5. This large anisotropy ratio provides evidence that the thermal transport along the inter-chain direction is significantly more resistive compared to the intra-chains.

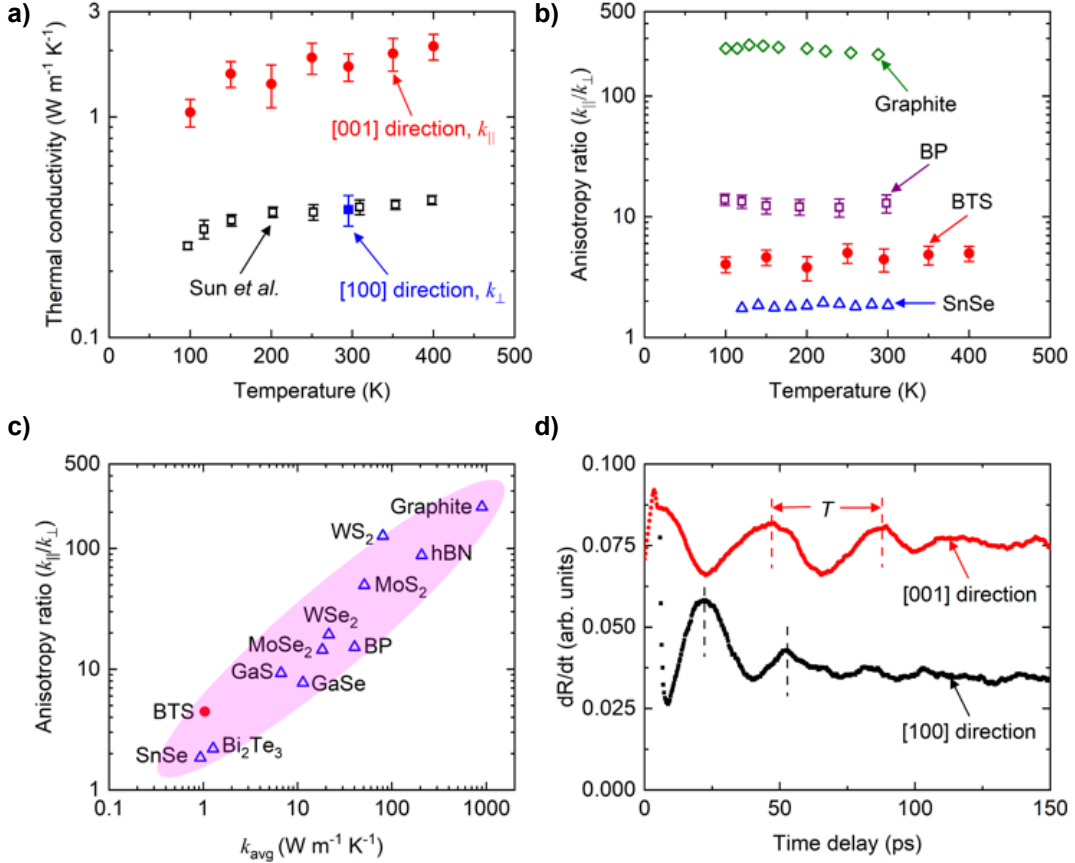

**Figure 4**. (a) Thermal conductivity of BTS as a function of temperature along the [001] and [100] directions. The temperature-dependent thermal conductivity along the [100] direction has been adopted from Sun et al.[17] (b) Anisotropy ratio as a function of temperature for BTS, graphite,[30] black phosphorus (BP),[31] and SnSe[32]. (c) Anisotropy ratio as a function of average thermal conductivity ($k_{avg} = \frac{k_{//} + k_{\perp}}{2}$) for BTS and other inorganic materials[30–32,34–39] with anisotropic crystal structure. The solid symbols represent measurements taken in this study and open symbols represent literature values. (d) Time periods ($T$) of the pressure front derivative along the [001] and [100] directions of the BTS crystal measured via TDBS. Here, R represents the ratio of in-phase to out-of-phase signal.

Moreover, temperature dependence of $k_{\parallel}$ follows the same glass-like thermal conductivity observed in the temperature dependence of $k_{\perp}$ (Figure 4a). Sun *et al.*[16] posited that the BTS thermal conductivity along the [100] direction exhibits an glass-like trend due to the enhanced phonon scattering mediated by tunneling of Ti ions between degenerate double well potential. To understand the nature of this thermal transport anisotropy, we compare the temperature-dependent anisotropy ratio of BTS along with that of graphite,[30] black phosphorus (BP),[31] and SnSe[32] in Figure 4b. In materials with anisotropic crystal structure such as graphite, BP, and SnSe, the thermal conductivity along different crystallographic directions is dictated by the same phonon scattering mechanism, hence the anisotropy ratio is nearly constant as a function of temperature.[32,33] Figure 4b shows that the anisotropy ratio of BTS is also nearly temperature independent. This indicates that the same phonon-scattering mechanism is likely dictating the thermal transport along the [001] and [100] directions.

In Figure 4c, we plot the thermal conductivity anisotropy ratio as a function of average thermal conductivity for inorganic materials having anisotropic crystal structures.[30–32,34–39] As demonstrated here, quasi-1D BTS is one of the most thermally insulating anisotropic crystalline material reported to date. In addition, its anisotropy ratio is significantly higher than that of other low thermal conductivity materials such as SnSe and $Bi_2Te_3$. This establishes BTS as a promising material for directional thermal regulation in applications needing thermal insulators.

We next focus on the microscopic origin of this anisotropic thermal transport in BTS. According to the kinetic theory, the lattice thermal conductivity of any crystalline material can be expressed as the following[39]

$$k = {1}/{3} C v^2 \tau \quad (1)$$

where $C$, $v$ and $\tau$ are the volumetric heat capacity, phonon group velocity, and phonon scattering times, respectively. Using equation (1) for the two crystallographic directions of BTS, we can obtain the following

$$\frac{k_\parallel}{k_\perp} = \left(\frac{v_\parallel}{v_\perp}\right)^2 \left(\frac{\tau_\parallel}{\tau_\perp}\right)$$

or

$$\left(\frac{v_\parallel}{v_\perp}\right)^2 \left(\frac{\tau_\parallel}{\tau_\perp}\right) = 4.5 \quad (2)$$

To determine $v_\parallel / v_\perp$, we measure the longitudinal sound speed ($v_L$) along the [001] and [100] directions, because the phonon group velocity scales with this parameter.[39] For this purpose, we use time-domain Brillouin scattering (TDBS), details of which can be found elsewhere.[40,41] Similar to TDTR, in this technique, a modulated pump laser beam launches a coherent acoustic wave that traverses through the surrounding media. The resulting pressure gradient related to the change in index of refraction causes the pulsed probe laser to be partially reflected. When the probe beam is incident on the sample surface at a normal angle, the period ($T$) of the pressure front derivative can be measured by TDBS and related to $v_L$ of the partially transparent media via

$$v_L = \lambda/(2rT) \quad (3)$$

where $\lambda$ is the probe beam wavelength and $r$ is the acoustic index of refraction. Details of these parameters associated with the BTS crystal have been provided in the Methods section.

Figure 4d shows the time periods ($T$) along the [001] and [100] directions of the BTS crystals measured via TDBS. From these measurements, using equation (3), we determine $v_L$ along the two directions to be $4.19 \pm 0.42$ and $4.62 \pm 0.42$ $km \cdot s^{-1}$, respectively. As $v_L$ is also an indicator of bond strength and stiffness,[42] our measurements reveal that the bond energy and stiffness along the [001] and [100] crystallographic orientations are nearly the same for the BTS crystal.

Within uncertainty $v_L$ is nearly the same along the two directions, and hence, $v_\parallel / v_\perp$ can also be reasonably approximated to have a value of 1. Therefore, equation (2) reduces to

$$\tau_\parallel/\tau_\perp \approx 4.5$$

This indicates that the phonon mean-free-path along the inter-chains is significantly shorter compared to the intra-chains. Thus, the anisotropy in BTS thermal transport originates from the differences in phonon scattering rate, not the phonon group velocity.

# VI. CONCLUSION

We have successfully tailored the crystal growth of BaTiS$_3$ (BTS) crystals to achieve different crystal orientations and morphologies. In addition to high-quality needle-like crystals and (100)-oriented platelet crystals, (001)-oriented platelet crystals were also synthesized. The out-of-plane orientation of the crystals was determined by high resolution XRD, and the in-plane orientation was determined by pole-figure analysis.

The comparison between (100)- and (001)-oriented BTS gives us thorough understanding of the local symmetry-induced electronic structure anisotropy. X-ray absorption spectroscopy (XAS) studies of Ti L$_{2,3}$ edges reveal large in-plane polarization linear dichroism in BTS (100) while BTS (001) is more isotropic. We observe additional multiplet absorption features between excitations corresponding to t$_{2g}$ and e$_g$ orbitals of Ti. The calculated electronic structure of BTS match well with the experimental XAS results. The crystal field splitting is anisotropic based on the XAS studies. We use the projected density of states (PDOS) from density functional theory (DFT) calculations for $d_{z^2}$, ($d_{x^2-y^2}$, $d_{xy}$) and ($d_{xz}$, $d_{yz}$) to explain the x-ray linear dichroism (XLD) behavior of BTS.

The mutually orthogonal (100)- and (001)-oriented BTS platelets complete the determination of full physical property tensor. BTS (001) crystal is optically isotropic in-plane, while BTS (100) shows giant in-plane optical anisotropy between the *a*-axis (ordinary axis) and the *c*-axis (extraordinary axis). BTS is a uniaxial positive birefringent crystal with the *c*-axis as the optical axis. In addition, BTS crystals are one of the most thermally insulating inorganic materials with substantial anisotropy (k$_∥$ / k$_⊥$ of ~4.5) reported to date. Different phonon scattering rates along the [001] and [100] directions give rise to this anisotropy.

Our study shows clear anisotropy in multiple physical properties of BTS, thoroughly characterized by the different orientations of the platelet crystals. These results reinforce the utility of BTS for next-generation imaging, communications, and sensing applications for miniaturized photonic devices, especially those requiring innovative thermal regulation.

# METHODS

## I. Crystal Growth

Single crystals of BTS were grown by chemical vapor transport with iodine as transporting agent. Starting materials, barium sulfide powder (Sigma-Aldrich, 99.9%), titanium powder (Alfa Aesar, 99.9%), sulfur pieces (Alfa Aesar, 99.999%) and iodine pieces (Alfa Aesar 99.99%) were stored and handled in a nitrogen-filled glove box. Stoichiometric quantities of precursor powders with a total weight of 1.0 g were mixed and loaded into a quartz tube of diameter 19 mm and thickness 2 mm along with around 0.75 $mg \cdot cm^{-3}$ iodine inside the glove box. The tube was capped with ultra-torr fittings and a bonnet needle valve to avoid exposure to the air. The tube was then evacuated and sealed using a blowtorch, with oxygen and natural gas as the combustion mixture. The sealed tube was about 12 cm in length and was loaded and heated in a Lindberg/Blue M Mini-Mite Tube Furnace or MTI OTF-1200X-S-II Dual Heating Zone 1200C compact split tube furnace.

In the synthesis, precursors were heated to reaction temperature at 100 °C/h and held for 100h before a slow cooling down to 950 °C at 10 °C/h. The furnace was then shut off letting the ampule to cool down within furnace. Dual zone furnace crystal growth used the same heating profile, while the temperature gradient was optimized to 8 °C/cm. Precursors were held at the center of the hot zone.

Crystals were manually collected by high precision tweezers under a Nikon SMZ660 Zoom Stereo Microscope equipped with a Micro Lite FL 1000 Ring Illuminator. High resolution morphology imaging was then carried out by JEOL 7000 FE scanning electron microscope and Olympus BX51 microscope equipped with an Infinity 2-1C CCD Digital Camera.

## II. X-ray Diffraction

XRD studies were performed in a Bruker D8 Advance X-ray diffractometer (XRD) in parallel beam configuration, using a germanium (004) two-bounce monochromator for Cu K$\alpha$1 ($\lambda$=1.5406 Å) radiation under a thin film out-of-plane configuration. Crystals are loaded on the Compact Cradle on top of a glass-slide holder. Rocking curve is achieved by varying the incident angle while remaining 2$\theta$ fixed at the center of the peak position. More details are in the Supporting Information II.

## III. Soft X-ray Absorption Spectroscopy (XAS)

Soft X-ray absorption spectroscopy was carried out at beamline 13–3 of the Stanford Synchrotron Radiation Lightsource (SSRL) using a transition edge sensor (TES) spectrometer. The TES spectrometer consists of a 240-channel energy-dispersive detector array facing the sample. The samples were mounted on a silicon substrate with double-sided carbon tape and transferred to the UHV chamber. The temperature was controlled by a liquid He filled cryostat. All spectra were normalized by the current from evaporated gold on a fine grid positioned upstream of the main chamber. Wavelength of incoming beam is selected by a Ni coated 600 l/mm Spherical Grating Monochrometer (SGM) with $TiO_2$ as reference. Incident photon polarizations of sigma (vertical linear), pi (horizontal linear), left-handed circular polarization, and right-handed circular polarization were selected by the elliptical polarized undulator (EPU), and each spectrum was collected 6 times for averaging.

The energy measured by the TES was calibrated using the theoretical fluorescence energies of $TiO_2$. Partial fluorescence yield (PFY) spectra of Ti were obtained by integrating the x-ray counts over the emission energy between 380 eV and 400 eV, which correspond to the 3$s$→2$p_{j=3/2}$ ($L_3$-edge) and 3$s$→2$p_{j=1/2}$ ($L_2$) florescence region of Ti. All spectra were normalized by the current from evaporated gold on a fine grid positioned upstream of the main chamber, which is roughly proportional to the incident beam flux. Data was analyzed in PyMca,[43] and peak fitting was done in Athena.[44]

## IV. Density Functional Theory (DFT) Calculations

All DFT calculations were performed within the generalized gradient approximation (GGA) with the Perdew-Burk-Ernzerhof (PBE) functional.[45] An energy cutoff of 650 eV and a $k$-point spacing of 0.03 Å$^{-1}$ were chosen to ensure that the total energy converged to 10$^{-6}$ eV/atom. Calculations of the Ti $L_{2,3}$ XAS spectra were performed using the CASTEP code.[46] For a specific Ti atom, a core-hole was included by removing a single electron from the Ti 2$p$ core states, which was treated by an excited state on-the-fly (OTF) pseudopotential. All other atoms were treated by standard ground-state OTF pseudopotentials. We used a 2×2×2 supercell to avoid fictitious interactions between the core-holes. To further analyze the orbital contribution in the XAS spectra, we calculated the orbital projected DOS using $Z + 1$ approximation,[47] which involved replacing the Ti atom ($Z = 22$) with a V atom ($Z = 23$). The $Z + 1$ calculations were performed with the Vienna Ab initio Simulation Package (VASP) code[48] using the same energy cutoff and $k$-point separation as the XANES calculation. All the calculated spectra were shifted along the energy axis to achieve the best fit to the experimental spectra.

## V. Fourier Transform Infrared Spectroscopy (FTIR)

Infrared spectroscopy was performed using a Fourier-transform infrared spectrometer (Bruker Vertex 70) connected to an infrared microscope (Hyperion 2000). A 15× Cassegrain microscope objective (numerical aperture = 0.4) was used for both transmission and reflection measurements at normal incidence. Polarization of the incident beam is controlled by a Thorlabs wire grid polarizer for 2-30 μm. FTIR measurements were performed with a Globar source, a potassium bromide beam splitter, and a mercury-cadmium-telluride (MCT) detector.

To perform the optical measurements, BTS (100) crystals were suspended from the edge of a Si handle wafer using Kapton tape. The pre-determined crystal orientation was manually matched with the polarizer orientation. Transmittance and reflectance spectra were collected every 15° within a 180° rotation for BTS (100). To present the transmittance and reflectance values at 2 μm, we averaged across one Fabry–Pérot fringe centered at 2 μm. BTS 001 crystals were mounted on a TEM grid over a hole size of 250 μm. The transmittance spectra were collected every 45°, and reflectance spectrum is collected every 22.5°.

## VI. Time-domain Thermoreflectance (TDTR)

TDTR setup has a Ti:sapphire laser emanating sub-picosecond laser pulses at a frequency of 80 MHz and central wavelength of 800 nm which is split into a high-power pump and a low-power probe path. An electro-optic modulator (EOM) then modulates the pump beam at a frequency of 8.8 MHz before it is frequency doubled to a wavelength of 400 nm. The incident time between the pump and probe beams at the sample surface is delayed via a mechanical delay stage. The periodic temperature oscillations created by the pump beam at the sample surface is detected by the time-delayed probe beam using a balanced photodetector and a lock-in amplifier.

The thermal properties are determined by fitting a radially symmetric, multi-layer heat diffusion model[18,49] to the ratio of in-phase to out-of-phase signal from the lock-in amplifier. The coaxially focused $1/e^2$ diameters of the pump and probe beams are ~16 and 10 μm, respectively. During the measurements, the power of the pump and probe lasers is adjusted to keep the steady-state temperature rise[50] below 6 K. All the BTS samples are coated with a 74 nm aluminum (Al) film via electron beam evaporation for optothermal transduction prior to TDTR measurements. The error bars incorporate the uncertainty of the Al thickness, Al thermal conductivity, Al and BTS heat capacity[33].

## VII. Time-domain Brillouin scattering (TDBS)

For the TDBS measurements, we use a two-tint version[51] of our TDTR setup, where both pump and probe wavelengths are 808 nm. Co-axially focused $1/e^2$ pump and probe diameters of ~4.5 μm are used for these measurements. The time-period of the pressure front derivatives along the [001] and [100] directions is measured to be 40 ± 4 and 33 ± 3 ps, respectively. The error bars of the time-periods represent the measurement repeatability over multiple peaks and multiple scans. The index of refraction along the two directions is measured by ellipsometry and has a value of 2.39 and 2.63, respectively.

# ACKNOWLEDGEMENTS

This work was supported by the Army Research Office (ARO) under award number W911NF-19-1-0137 and via an ARO MURI program with award number W911NF-21-1-0327, the National Science Foundation of the United States under grant numbers DMR-2122070 and 2122071, and the USC Provost New Strategic Directions for Research Award. M.S.B.H., M.M., J.A.T., and P.E.H. acknowledge support from the Army Research Office under award number W911NF-21-1-0119. H.M and M.K acknowledge the support from the Office of Naval Research (N00014-20-1-2297), N.W and H.W acknowledge the support by the NSF FMRG program (CMMI-2036359). The soft x-ray spectroscopy experiments were carried out at the SSRL (beamline 13-3), SLAC National Accelerator Laboratory, supported by the U.S. Department of Energy, Office of Science, Office of Basic Energy Sciences under contract no. DE-AC02-76SF00515. The authors gratefully acknowledge the use of facilities at Dr. Stephen Cronin's Lab, and Core Center for Excellence in Nano Imaging at University of Southern California for the results reported in this manuscript. B.Z acknowledges technical assistance from Jieyang Zhou, Mythili Surendran, Thomas Orvis and Harish Kumarasubramanian in collaboration in the related projects.

# Supporting Information

## I. Structure configurations of BTS crystals

Figure S1 shows a) BTS needle, b) BTS (100) and c) BTS (001). The orientation determined by XRD are also plotted next to the crystal. Optical images of BTS crystals are obtained using an Olympus BX51 microscope.

BTS needle takes the form of a one-dimensional needle. Its *c*-axis is parallel to the long axis (along the length of the needle), and *a*-axis and *b*-axis usually perpendicular to the length. A regular shaped BTS (100) has a rectangular morphology with the edges parallel either to *b*-axis or *c*-axis, making the third axis (*a*-axis here) 60 degrees out-of-plane. BTS (100) usually has terraces on the surface. BTS (001) has a hexagonal morphology. The edges are usually parallel to *a*-axis or *b*-axis. Terraces and islands easily form on the surface of BTS (001) with hexagonal morphology similar to smaller platelets. A very rough surface is therefore optically visible.

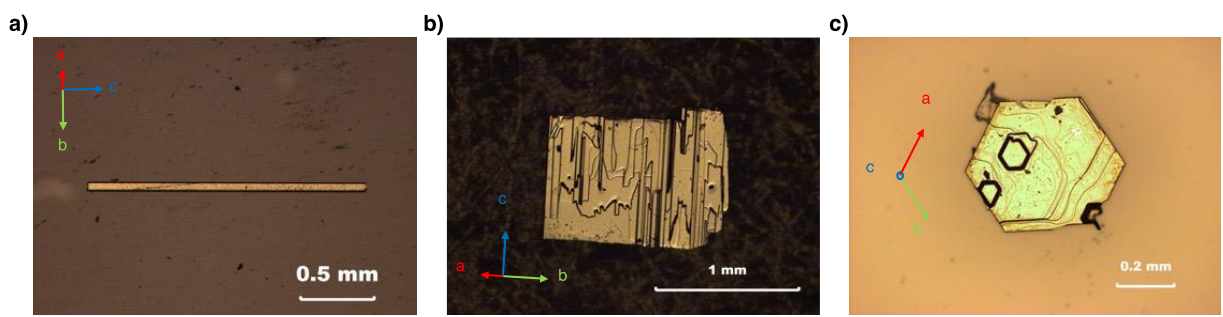

**Figure S1**. Optical images and crystal orientations of BTS crystals of three morphologies. (a) BTS needle; (b) BTS (100); (c) BTS (001).

## II. Out-of-plane and in-plane orientation by thin film XRD

The orientation of the terminating surfaces of BTS are characterized by X-ray diffractometer (XRD) based on the presumed crystalline orientations. Figure S2 c, d, e shows the out-of-plane XRD of BTS needle, BTS (100) and BTS (001), and the insets are the rocking curves for 300 or 002 reflections. The rocking curves are fitted with a gaussian distribution to extract the full width at half maximum (FWHM). XRD of BTS needle and BTS (100) points to the same out-of-plane orientation, thus both structures are terminated by {100} crystalline surfaces. However, the rocking curve of BTS (100) doesn't have a perfect gaussian distribution even if the FWHM of 0.011° is lower than that of BTS needle. This higher mosaicity in BTS (100) plate could originate from small angle domain boundaries inside the crystal, while such domain rarely exists in BTS needles unless it is curved. BTS (001), on the other hand, are easily distinguished from BTS (100) by the difference in the number of 00*l* reflections: Although 00*l* and 0*k*0 reflections are only slightly different in the *d*-spacing (2θ), even numbers of 00*l* reflections of BTS are extinct given the sliding symmetry in *P*6$_3$*mc*. The crystal quality of BTS (001) is comparable with other morphologies with a rocking curve of 0.013° FWHM.

The in-plane crystallography orientation of BTS crystals and facets are determined by pole figure studies. Figure S2a, b shows the geometry within a qualitative Ewald's sphere above the sample stage. The crystal is at the center of the hemisphere, while the source and detector are located on the hemisphere and can be moved independently along a longitude circle and thus such plane is the diffraction plane and 2θ is the angle between them. ζ represents the angle between the out-of-plane orientation and the bisector of 2θ. The ϕ is the rotational axis of the goniometer, perpendicular to the sample stage and the χ is the tilting axis of the goniometer, representing the tilting angle between the rotational axis and the diffraction plane. Out-of-plane orientation will then be defined by χ = 0 in this geometry. Although tilting of 90° can be achieved with χ, samples with platelet geometry lose diffraction intensity dramatically at high tilting angle due to reduced beam projection. Therefore, off-axis reflections close to the out-of-plane are best for in-plane orientation determination.

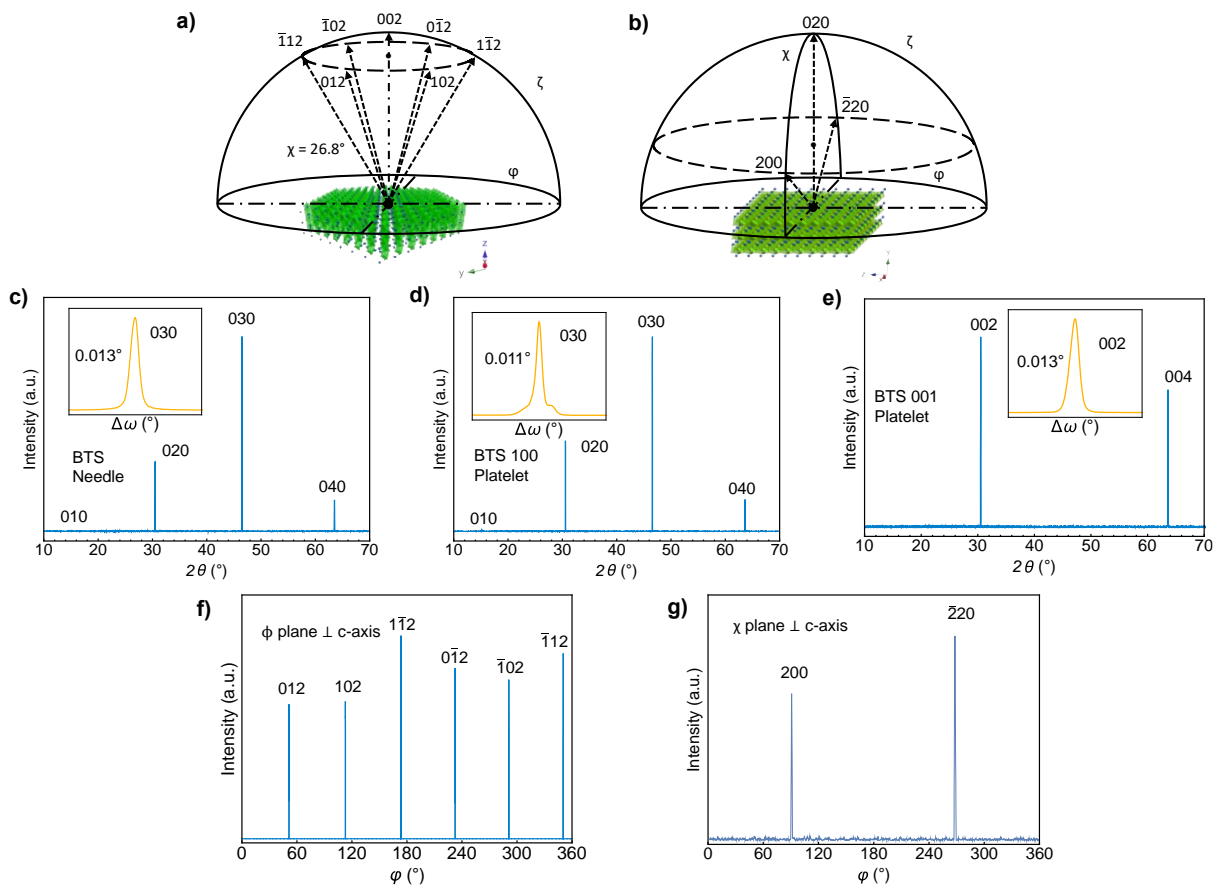

**Figure S2.** Orientation determination of (100) and (001) BTS plates. (a) XRD geometry for BTS (001). 002 reflects the out-of-plane orientation. 6-fold symmetry is determined by pole figure. Angle between 002 and 012, χ=26.8°, is calculated form the crystal structure; (b) XRD geometry for BTS needle and BTS (100). The out-of-plane orientation is (001). 6-fold symmetry is determined by 60° χ-rotation between 200 and 020 orientations. Thin film out-of-plane XRD and the rocking curve of (c) BTS needle, (d) BTS (001), (e) BTS (100). BTS needle and BTS (100) have the same 0$k$0 orientation and a full width at half maximum (FWHM) of 0.013° and 0.011° in the rocking curve. BTS (100) shows more mosaicity, represented by the asymmetric rocking curve. BTS (001) has a 00$l$ orientation, and a rocking curve with FWHM of 0.013°. (f) Pole figure of BTS (001) crystal. 012 off-axis reflections show 6-fold rotational axis perpendicular to ϕ. The asymmetry in the reflection comes from the misalignment; (g) Pole figure of BTS (100) crystal. χ = 60° between 020 and 200 matches the 6-fold symmetry. The in-plane orientation is also determined by a pole figure, while $c$-axis is in the diffraction plane when 200 or $\bar{2}$20 reflection appear.

BTS (001) has 012 off-axis reflections at 26.8° from 002 and shows a six-fold rotational symmetry about 002. To find such reflections, BTS (001) is first aligned towards out-of-plane orientation like we did earlier. Then a tilting angle χ or ζ is set to 26.8° which is carefully examined based on the reference crystal structure. Finally, we set 2θ = 34.35° for 012 reflections (or 2θ angle for 024 if ζ tilting is used) and rotate the crystal with ϕ from 0° to 360°. This will create the pole figure for BTS (001), and six equivalent reflections given by six-fold symmetry are detected if the crystal is properly aligned (Figure 2f). 100 is therefore lying in the diffraction plane when we are at the center of 012 reflections. Similar method is used for BTS (100) and BTS needle to determine $c$- and $a$-axis. However, one needs to be careful as in-plane reflections ($4\bar{2}2$, $4\bar{2}\bar{2}$, $\bar{4}22$, $\bar{4}2\bar{2}$ and 004, $00\bar{4}$ for example) are almost indistinguishable with 2θ angles lying about 60° away in {100} planes. This will mess up the symmetry determination. We use the primary reflections of $h00$ to tell the in-plane orientations apart. Figure 2g shows the diffraction geometry of the in-plane orientation determination of BTS (100), and the pole figure is shown in Figure 2b. The $c$-axis will be within the diffraction plane when 200 or $\bar{2}20$ reflection appears.

### III. Pole figure analysis

To exclude incommensurate superlattices along the primary axes, rotational XRD ζ mapping are done in $a$-$b$ and $a$-$c$ plane. This is a series of θ-2θ scans with stepped ζ angle. Once the $c$-axis is aligned perfectly parallel to the out-of-plane orientation, a ζ mapping will demonstrate all reflections between 100 and 010 in the $a$-$b$ plane (Figure S3a) or $a$-$c$ plane if the $a$-axis is parallel to the out-of-plane (Figure S3b). In Figure S3a, 010 orientation is first aligned to the out-of-plane orientation, then rotation b-axis parallel to χ rotational plane. θ-2θ scans from 10° to 60° are carried out for every 1° between χ = -6° to 66°. Peaks are indexed based on 2θ and χ relative to out-of-plane orientation (100 series in this case) and plotted in 2θ space. This verified the six-fold axis to be perpendicular to χ rotational plane. Similar mapping is done for a-c plane for χ = -5° to 75°, Figure S3b. No incommensurate structures are seen in $a$-$b$ and $a$-$c$ plane. This matches with the reported crystal structure.

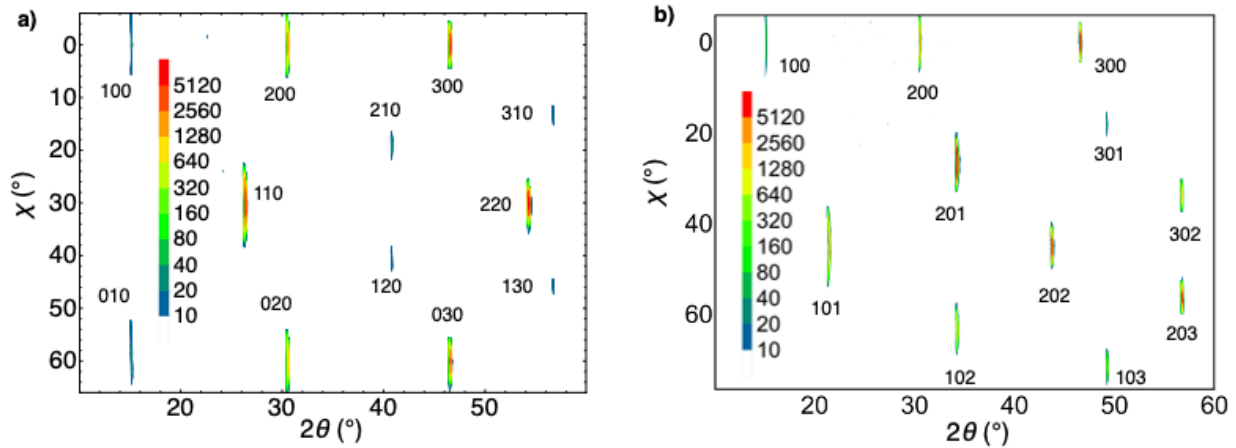

**Figure S3**. Rotational XRD Mapping (θ-2θ scan and map χ), and sign of superstructure. (a) a-b plane. A 60° rotation from 100 to 010 is characterized. Peaks showing up match with diffraction pattern of BTS; (b) a-c plane. No incommensurate superstructure shows up along a-axis or c-axis. Structure matches with a stoichiometry BaTiS$_3$.

### IV. X-ray absorption spectrometry (XAS) in total electron yield (TEY)

Total electron yield is also monitored in XAS measurements as the compensating current flowing into the sample. Figure. S4 a, b show the polarization averaged XAS spectra of BTS (100) and BTS (001) in TEY mode and PFY mode. Because the average electron mean free path at 460 eV is around 1.2 nm[52] while the photon mean free path at 460 eV is estimated as 0.2 μm[44]. TEY and PFY XAS is then sensitive to different thickness of BTS crystal: TEY is detecting the first few layers of atoms near the surface while PFY is more sensitive to the bulk down to 0.2 µm. This comparison indicates the surface states of the as grown BTS crystals are different from the bulk towards higher XAS energy. A reasonable guess is the oxygen absorbance at the surface has stabilizes oxidized states and passivated the surface. We then used PFY mode for BTS XAS analysis to minimize surface states contribution.

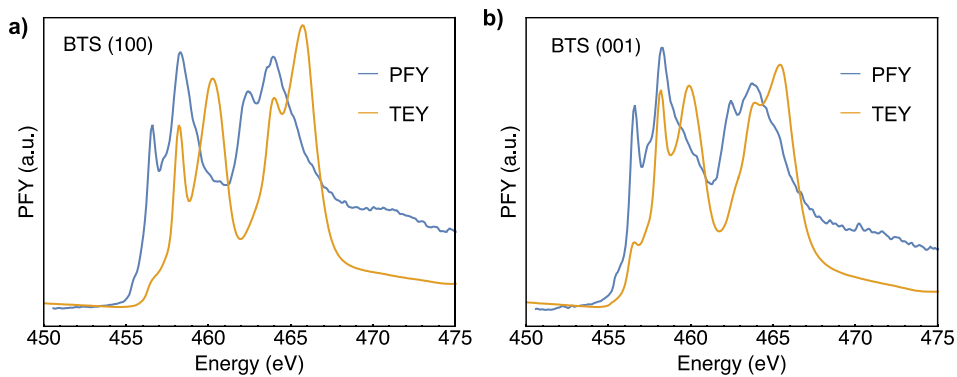

**Figure S4**. Polarization averaged XAS spectra of (a) BTS (100), and (b) BTS (001) in PFY mode and TEY mode measured simultaneously. TEY shows surface states of higher energy, which is most likely to be oxidized states of Ti near the surface.

### V. Additional measurements using Fourier-transform infrared (FTIR) spectroscopy

Unpolarized transmittance spectra of BTS crystals with different morphologies grown at different times were measured in FTIR. BTS crystals were mounted on top of the same sapphire substrate. A blank substrate was then used as the background. Spectra of different crystals were scaled to make a clear comparison in Figure S5. Except for BTS (001), BTS crystals all features two absorption edges at 0.2-0.3 eV and 0.75-0.8 eV. Morphologies and small temperature change within the growth window thus do not change the optical properties of BTS.

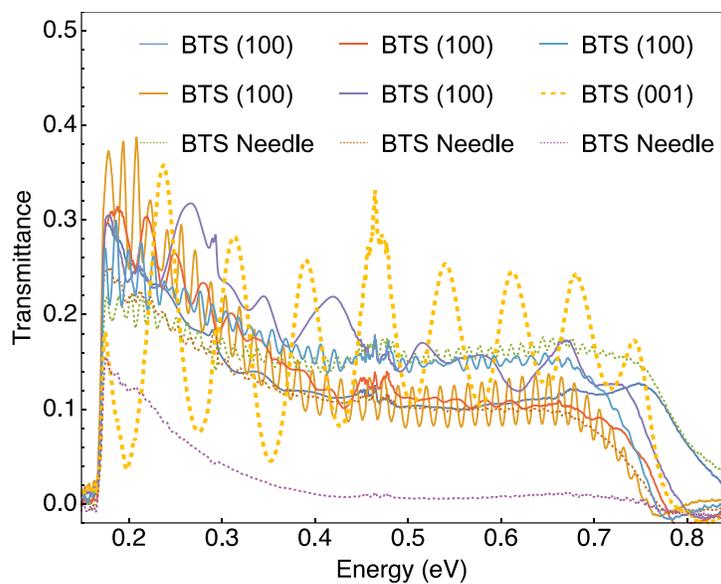

**Figure S5**. FTIR transmittance spectra of BTS crystals of different morphologies. Two absorption energies exist for BTS needle and BTS (100): 0.76 eV belonging to ordinary (⊥ c) and 0.27 eV which is extraordinary (∥ c). BTS (001) does not show extraordinary transmittance and have only one absorptance edge at 0.76 eV.